\documentclass{IEEEtran4PSCC}
\ifCLASSINFOpdf
\else
\fi
\usepackage{amsmath,amssymb}
\usepackage{enumitem}
\usepackage[linktocpage=true,colorlinks=true,linkcolor=blue,citecolor=blue,urlcolor=blue]{hyperref}
\usepackage{graphicx} % Required for inserting images
\usepackage{multirow}
\usepackage{listings}

\usepackage{inconsolata} % very nice fixed-width font included with texlive-full
\usepackage[usenames,dvipsnames]{xcolor} % more flexible names for syntax highlighting colors 
\usepackage{colortbl}
\usepackage{caption}
\usepackage{cite}
\lstdefinelanguage{julia}
{
  keywordsprefix=\@,
  morekeywords={
    exit,whos,edit,load,is,isa,isequal,typeof,tuple,ntuple,uid,hash,finalizer,convert,promote,
    subtype,typemin,typemax,realmin,realmax,sizeof,eps,promote_type,method_exists,applicable,
    invoke,dlopen,dlsym,system,error,throw,assert,new,Inf,Nan,pi,im,begin,while,for,in,return,
    break,continue,macro,quote,let,if,elseif,else,try,catch,end,bitstype,ccall,do,using,module,
    import,export,importall,baremodule,immutable,local,global,const,Bool,Int,Int8,Int16,Int32,
    Int64,Uint,Uint8,Uint16,Uint32,Uint64,Float32,Float64,Complex64,Complex128,Any,Nothing,None,
    function,type,typealias,abstract
  },
  sensitive=true,
   alsoother={$},%
   morecomment=[l]\#,%
   morecomment=[n]{\#=}{=\#},%
   morestring=[s]{"}{"},%
   morestring=[m]{'}{'},%
}[keywords,comments,strings]%

\usepackage{svg} 
\newcommand{\ie}{\emph{i.e.}}

\newcommand{\st}{\mathop{\text{\normalfont s.t.}}}

\usepackage[nameinlink]{cleveref}
\crefname{equation}{}{}
\Crefname{equation}{}{}
\crefname{assumption}{Ass.}{Ass.}
\Crefname{assumption}{Assumption}{Assumptions}
\creflabelformat{equation}{(#2#1#3)}

\usepackage{pgfplots, pgfplotstable}
\usepackage{glossaries}
\newacronym{ac}{AC}{alternating current}
\newacronym{ad}{AD}{automatic differentiation}
\newacronym{dc}{DC}{direct current}
\newacronym{opf}{OPF}{optimal power flow}
\newacronym{gpu}{GPU}{graphics processing unit}
\newacronym{mp}{MP}{multi-period}
\newacronym{sc}{SC}{security-constrained}
\newacronym{goc3}{GOC3}{Grid Optimization Competition Challenge 3}
\newacronym{tso}{TSO}{transmission system operators}
\newacronym{so}{SO}{system operators}
\newacronym{nlp}{NLP}{nonlinear programming}
\newacronym{ml}{ML}{machine learning}
\pgfplotsset{compat=1.18}
\usetikzlibrary{patterns}

\makeatletter
\let\old@ps@headings\ps@headings
\let\old@ps@IEEEtitlepagestyle\ps@IEEEtitlepagestyle
\def\psccfooter#1{%
    \def\ps@headings{%
        \old@ps@headings%
        \def\@oddfoot{\strut\hfill#1\hfill\strut}%
        \def\@evenfoot{\strut\hfill#1\hfill\strut}%
    }%
    \def\ps@IEEEtitlepagestyle{%
        \old@ps@IEEEtitlepagestyle%
        \def\@oddfoot{\strut\hfill#1\hfill\strut}%
        \def\@evenfoot{\strut\hfill#1\hfill\strut}%
    }%
    \ps@headings%
}
\makeatother

\psccfooter{%
        \parbox{\textwidth}{\hrulefill \\ \small{24th Power Systems Computation Conference} \hfill \begin{minipage}{0.2\textwidth}\centering \vspace*{4pt} \includegraphics[scale=0.06]{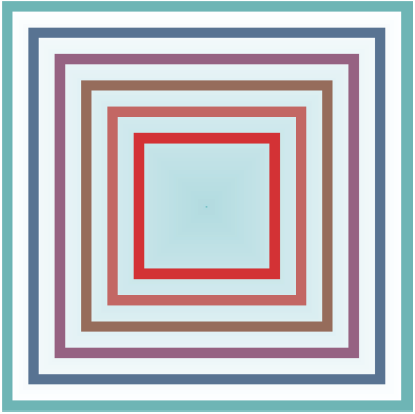}\\\small{PSCC 2026} \end{minipage} \hfill \small{Limassol, Cyprus --- June 8-12, 2026}}%
}

\begin{document}
\title{ExaModelsPower.jl: A GPU-Compatible Modeling Library for Nonlinear Power System Optimization}

\author{
\IEEEauthorblockN{Sanjay Johnson, Dirk Lauinger, Sungho Shin}
\IEEEauthorblockA{Massachusetts Institute of Technology \\
Cambridge, Massachusetts, USA\\
\{sanjayjo, lauinger, sushin\}@mit.edu}
\and
\IEEEauthorblockN{François Pacaud}
\IEEEauthorblockA{
Mines Paris-PSL \\
Paris, France \\
francois.pacaud@minesparis.psl.eu}
}

\maketitle

\begin{abstract}
  As GPU-accelerated mathematical programming techniques mature, there is growing interest in utilizing them to address the computational challenges of power system optimization.
  This paper introduces ExaModelsPower.jl, an open-source modeling library for creating GPU-compatible nonlinear AC optimal power flow models.
  Built on ExaModels.jl, ExaModelsPower.jl provides a high-level interface that automatically generates all necessary callback functions for GPU solvers.
  The library is designed for large-scale problem instances, which may include multiple time periods and security constraints.
  Using ExaModelsPower.jl, we benchmark GPU and CPU solvers on open-source test cases. 
  Our results show that GPU solvers can deliver up to two orders of magnitude speedups compared to alternative tools on CPU for problems with more than 20,000 variables and a solution precision of up to $10^{-4}$, while performance for smaller instances or tighter tolerances may vary.
\end{abstract}


\section{Introduction}\label{sec:intro}

\Gls*{ac} \gls*{opf} is used to determine optimal power system operating points, which must satisfy nonlinear power flow constraints and various operational constraints~\cite{cain_history_2012,frank_introduction_2016,coffrin_powermodels_2018}.
Resolving \gls*{ac}\gls*{opf} problems, formulated as \gls*{nlp}, can be computationally challenging, especially for large networks with thousands of buses, lines, and generators.
The development of efficient solution methods for these problems remains an active area of research.
In practice, \gls*{so} often solve \gls*{dc} \gls*{opf} problems, \ie, linear approximations that neglect the inherent nonlinearities of the power flow equations~\cite{cain_history_2012, oneill_recent_2011}.
Such approximations can lead to suboptimal or infeasible solutions as \gls*{dc} feasibility does not guarantee \gls*{ac} feasibility.
It is estimated that switching from \gls*{dc} to \gls*{ac}\gls*{opf} could reduce electricity costs by several percentage points, translating into billions of dollars in yearly savings for the US~\cite{cain_history_2012}. Modeling with\gls*{ac}\gls*{opf} not only leads to efficiency improvements but also improves reliability by considering the roles of reactive power and voltage magnitude. 
Numerous efforts, such as the ARPA-E Grid Optimization Competitions~\cite{aravena_recent_2023,holzer_grid_2024,holzer2025go, elbert2024final_report}, have thus been undertaken to improve solution methods for \gls*{ac}\gls*{opf}. Results from the competition show that solving the nonlinear power flow equations remains a computational bottleneck for large-scale systems~\cite{brun2025alternating, chevalier2024parallelized, sharadga2025scalable}.

The scale of \gls*{ac}\gls*{opf} depends on the size of the underlying physical network, the number of time periods, and the number of security constraints. \Gls*{mp} problems are necessary to capture the dynamics of energy storage, flexible loads, and ramp-rate-constrained generators~\cite{geth_flexible_2020}. \Gls*{sc} problems increase system resiliency by requiring that operating points be able to withstand contingencies, \ie, individual component failures. The \gls*{goc3} includes test cases with up to 48~periods and thousands of contingencies~\cite{petra_solving_2021, holzer_grid_2024}. These test cases were designed to be realistic and pose significant computational challenges.

Solution methods for large-scale \gls*{ac}\gls*{opf} would ideally exploit the sparsity of the underlying physical network. This was initially difficult for \glspl*{gpu}, due to challenges with sparse automatic differentiation (AD) and sparse linear solver routines. These issues have recently been addressed by single-instruction multiple-data (SIMD) abstractions of nonlinear programs and condensed-space interior-point methods, respectively, enabling order-of-magnitude speedups, though challenges remain in obtaining high-precision solutions~\cite{shin_accelerating_2024}.

In order to efficiently run these \gls*{nlp} solvers on \gls*{gpu}, a modeling library must be able to provide objective, constraint, Jacobian, and Hessian evaluation methods via \gls*{gpu}-compatible callback functions.
Specifically, all the array data needs to stay within the \gls*{gpu} device memory, and all the callback functions need to be implemented as \gls*{gpu} kernels without any data transfer between the host and device memory; otherwise, problem data would be transferred between host memory and \gls*{gpu} device memory at each iteration, causing serious overhead.
On the contrary, implementing custom callback functions as \gls*{gpu} kernels is a cumbersome and error-prone task.
This motivates the development of a modeling library that automatically generates all necessary callback functions from a high-level user interface.

In this paper, we introduce such a library, ExaModelsPower.jl. This open-source power system modeling library
employs ExaModels.jl~\cite{shin_accelerating_2024} as an abstractions layer to create \gls*{gpu}-compatible \gls*{nlp} problems.
ExaModels.jl automatically generates \gls*{gpu} kernels for model evaluation and differentiation from the objective and constraints specified within ExaModelsPower.jl, which are expressed as human-readable algebraic expressions.
This implementation enables GPU-compatible power system models to be created via a high-level user interface (similar to PowerModels.jl~\cite{coffrin_powermodels_2018} or MATPOWER~\cite{zimmerman_matpower_2011}). 
The model created with ExaModelsPower.jl can be evaluated on heterogeneous hardware backends, including both CPU and \gls*{gpu}, and can be solved with a variety of solvers (any solver available within NLPModels.jl ecosystem  \cite{orban_nlpmodelsjl_2023}).

\paragraph*{\textbf{Survey of Existing Power System Modeling and Optimization Libraries}}
There are several layers in the power-system decision-making process, each applied at different time scales.
Common practices include capacity expansion planning (months to years), unit commitment (hours to days), and \gls*{opf} (minutes to hours).
Dynamics at faster time scales (seconds to minutes) are typically handled by specialized control systems rather than mathematical optimization methods.

Capacity expansion planning determines  investments in generation, transmission, and storage assets over multi-year horizons. 
Software tools for capacity expansion planning include GenX~\cite{bonaldo_genx_2025}, ReEDS~\cite{ho_regional_2021}, and PyPSA~\cite{brown_pypsa_2018}.

Unit commitment determines the scheduling of generation units, which typically involves logical constraints modeled by auxiliary binary variables. 
Software tools for unit commitment include PLEXOS~\cite{c_papadopoulos_r_johnson_and_f_valdebenito_plexos_integrated_plexos_2014} and UnitCommitment~\cite{santos_xavier_unitcommitmentjl_2024}.

\gls*{ac}\gls*{opf} determines power system operating points that respect all operational constraints, including the nonlinear power flow equations, at least cost.
PowerModels.jl and MATPOWER provide diverse options for user-specified \gls*{ac}\gls*{opf} models, respectively in Julia and Matlab~\cite{coffrin_powermodels_2018,zimmerman_matpower_2011}. PowerModels.jl uses JuMP~\cite{lubin_jump_2023} as a modeling backend, while MATPOWER implements its own custom derivative evaluation. PYPOWER ports MATPOWER to allow for \gls*{opf} modeling in Python~\cite{lincoln_pypower_2025}. Pandapower is built on pandas---a data analysis library in Python---and provides a Python-based tool for power system modeling and analysis~\cite{thurner_pandapoweropen-source_2018}.
Pandapower can be used for optimization, but relies on PowerModels.jl or PYPOWER as external modeling backends. ExaGo provides the most similar functionality to ExaModelsPower.jl, as it also aims to provide an interface for creating \gls*{gpu}-compatible \gls*{ac}\gls*{opf} models~\cite{noauthor_exago_nodate}. ExaGo uses~PETSc~\cite{balay_petsctao_2025} as modeling backend with RAJA~\cite{beckingsale_raja_2019} and Umpire~\cite{beckingsale_umpire_2020} for GPU memory management and loop abstractions.
To our knowledge, in their current version v1.6.0, \gls*{ac}\gls*{opf} models cannot be solved in a fully \gls*{gpu}-resident manner because none of the solvers interfaced to ExaGo are fully \gls*{gpu}-compatible.
For example, HiOp currently handles mixed sparse-dense formulations and does not support the fully sparse formulations required for large-scale \gls*{ac}\gls*{opf} problems \cite{petra_hiop_2018}.
We are not aware of any published performance benchmarking of ExaGo running on \gls*{gpu}.

In addition to optimization methods, there is growing interest in training \gls*{ml} proxies for \gls*{opf} solvers.
Various strategies have emerged, which have the potential to greatly reduce online computation time by shifting the computational burden to an offline training phase~\cite{khaloie2025review,donti2021dc3, fioretto2020predicting, pilotoCANOSFastScalable2024}.
However, handling various operational constraints and ensuring feasibility over various operating conditions remain significant challenges.
As of now, these methods have been mostly demonstrated on a hand-curated, limited number of problem instances, and to the best of our knowledge, none of these methods have been demonstrated over a large datasets, such as the entire pglib-opf library~\cite{babaeinejadsarookolaee_power_2021}.

\paragraph*{\textbf{Contributions}}
This paper makes two main contributions.
First, ExaModelsPower.jl is introduced as a power system modeling library that provides the capability to automatically construct the \gls*{gpu} kernels for model evaluation and differentiation.
A number of \gls*{opf} modeling platforms exist, but none of the existing tools enable solving \gls*{ac}\gls*{opf} models entirely on \gls*{gpu} as ExaModelsPower.jl does.
This paper provides various code snippets to demonstrate the core functionalities of ExaModelsPower.jl, ranging from simple static \gls*{ac}\gls*{opf} to more complex \gls*{mp} and \gls*{sc}\gls*{opf} models.
Furthermore, we demonstrate how users can extend existing models to incorporate custom model components.
Second, comprehensive benchmark results, including \gls*{gpu} and CPU solvers, are presented for a range of model variations.
While static and \gls*{mp} \gls*{opf} models have been previously benchmarked \cite{shin_accelerating_2024,shin_scalable_2024}, this paper provides a more comprehensive benchmark by including multiple solvers, model formulations, and a broader set of test cases.
In particular, we present benchmarking results for the \gls*{mp}\gls*{opf} with storage and the \gls*{goc3}-variant of \gls*{sc}\gls*{opf} formulation \cite{holzer_grid_2024}.

The paper proceeds as follows: \Cref{sec:simd} introduces ExaModels.jl, which is the underlying platform on which ExaModelsPower.jl is implemented. This section discusses how GPU compatibility is achieved at the modeling level.
\Cref{sec:exap} provides an overview of the various models implemented in ExaModelsPower.jl along with examples.
\Cref{sec:num} provides comprehensive benchmarking results for both \gls*{gpu} and CPU solvers on the models currently available in ExaModelsPower.jl. \Cref{sec:conc} provides several concluding remarks.
This paper does not aim to provide a comprehensive survey of \gls*{gpu}-based optimization. To learn more about \gls*{nlp} algorithms on \gls*{gpu}s, we refer the reader to \cite{shin_accelerating_2024,petra_hiop_2018,regev_hykkt_2023}.

%%% Local Variables:
%%% mode: LaTeX
%%% TeX-master: "main"
%%% End:

\section{Overview of ExaModels.jl and GPU-Compatible Nonlinear Optimization Modeling}\label{sec:simd}
This section provides an overview of ExaModels.jl, a platform over which ExaModelsPower.jl is implemented, and discusses how \gls*{gpu}-compatible \gls*{nlp} models can be created.
This section presents the design principles of ExaModels.jl, its abstraction---SIMD abstraction of \gls*{nlp}---, and its modeling syntax.
A more detailed description can be found in \cite{shin_accelerating_2024}.

\subsection{Core Features of ExaModels.jl}
ExaModels.jl is a general-purpose modeling interface for solving \glspl*{nlp} of the following form:
\begin{equation*}
\label{eqn:nlp}
  \min_{x^\flat\leq x \leq x^\sharp} \, f(x) ~~ \st \, g^\flat \leq g(x) \leq g^\sharp,
\end{equation*}
where $x\in\mathbb{R}^n$ is the decision variable, $f:\mathbb{R}^n\to\mathbb{R}$ is the objective function, $c:\mathbb{R}^n\to\mathbb{R}^m$ is the constraint function, and $x^\flat,x^\sharp\in\mathbb{R}^n$ and $c^\flat,c^\sharp\in\mathbb{R}^m$ are the lower and upper bounds on the decision variables and constraints, respectively.

\Gls*{nlp} software toolchain consists of three main modular components: a modeling interface---such as JuMP \cite{lubin_jump_2023}, Pyomo \cite{hartPyomoModelingSolving2011}, CasADi \cite{anderssonCasADiSoftwareFramework2019}, AMPL Solver Library \cite{fourerAMPLMathematicalProgramming}, a solver---such as Ipopt \cite{wachter_implementation_2006}, KNITRO \cite{pardalos_knitro_2006}---, and a linear algebra backend---such as HSL library \cite{duffMA57aCodeSolution2004} and Pardiso \cite{schenkSolvingUnsymmetricSparse2004}.

ExaModels.jl is a modeling interface implemented in Julia Language \cite{bezansonJuliaFreshApproach2017}.
The modeling interface is a front-end for the users to specify the \gls*{nlp} model, including decision variables, objective function, and constraints, and provides callback functions for solvers to evaluate and differentiate the model equations.
As such, the main feature of ExaModels.jl is to provide a high-level user interface for specifying NLP models while automatically generating the callback functions.
Specifically, ExaModels.jl provides these callback functions in such a way that they can be executed on various hardware backends, including CPU (single or multi-threaded) and \gls*{gpu} architectures (e.g., NVIDIA, AMD and Intel).

\subsection{NLPModels.jl Abstraction}
ExaModels.jl creates \gls*{nlp} model that conforms to the NLPModels.jl template \cite{orban_nlpmodelsjl_2023}.
NLPModels.jl provides a standardized template for \gls*{nlp} problems.
In particular, it defines a set of templates for callback functions that serve as the common interface between modeling interfaces and solvers.
By conforming to NLPModels.jl template, the created model can become automatically compatible with any solver that is available in NLPModels.jl ecosystem, including MadNLP.jl \cite{shin_accelerating_2024}, Ipopt \cite{wachter_implementation_2006}, and KNITRO \cite{pardalos_knitro_2006}.
The callback functions required by NLPModels.jl are described in \Cref{nlpmodels}.
\begin{lstlisting}[language=Julia, caption={NLPModels callbacks}, label={nlpmodels}]
# returns objective function at x
NLPModels.obj(model, x)
# evaluates objective gradient at x in place of g
NLPModels.grad!(model, x, g)
# evaluates constraint function at x in place of c
NLPModels.cons!(model, x, c)
# evaluates constraint Jacobian at x in place of J
NLPModels.jac!(model, x, J)
# evaluates Lagrangian Hessian at x and l in place of H
NLPModels.hess!(model, x, l, H; obj_weight) 
\end{lstlisting}
We emphasize that 
NLPModels.jl itself is not a modeling interface, and is simply serving as a standard template for \glspl*{nlp}.
ExaModels.jl creates callback functions in \Cref{nlpmodels} based on the user-specified algebraic expressions.

\subsection{SIMD Abstraction}

The core design principle of ExaModels.jl is to maintain the model equations in a form that is amenable to \emph{Single-instruction-multiple-data (SIMD) parallelism}, which ensures that the structure within the model is exploited to the fullest extent. We refer to this design principle as SIMD abstraction, which presumes that the model equations can be expressed in the following form:
\begin{align}
\label{eqn:prob}
\min_{x^\flat\leq x \leq x^\sharp}\; & {\sum_{l=1}^{L}\sum_{i=1}^{I_l} f^{(l)}(x; p^{(l)}_i)} \\
\st\; & g^\flat_m \leq \sum_{n=1}^{N_m}\sum_{k=1}^{K_n}g^{(n)}(x; q^{(n)}_{k}) \leq g^\sharp_m, & \forall m=1,\cdots,M,\nonumber
\end{align}
where $f^{(\ell)}(\cdot,\cdot)$ and $g^{(m)}(\cdot,\cdot)$ are the components of the objective and constraint functions, and $\{\{p^{(k)}_i\}_{i\in [N_k]}\}_{k\in[K]}$, $\{\{\{q^{(n)}_{k}\}_{k\in[K_n]}\}_{n\in[N_m]}\}_{m=1}^M$ are
problem data, which can either be discrete or continuous.

One can observe that the objective functions and constraints are expressed as a collection of a common computational pattern, $f^{(l)}(\cdot,\cdot)$ and $g^{(n)}(\cdot,\cdot)$, respectively, repeatedly applied over multiple data entries.
Evaluating and differentiating each term can be performed independently, which enables the application of SIMD parallelism.
By designing modeling syntax and data structures to maintain the SIMD abstraction as in \cref{eqn:prob}, ExaModels.jl is equipped with structured data over which highly efficient callback functions can be created in the form of \gls*{gpu} kernels.
In particular, ExaModels.jl applies reverse-mode \gls*{ad} to each term in \eqref{eqn:prob} independently, and aggregates the results to construct the gradient, Jacobian, and Hessian of the entire model, with automatic sparsity detection.
More details on how the \gls*{ad} is implemented over the SIMD abstraction can be found in \cite{shin_accelerating_2024}.

\subsection{Modeling Syntax}

To implement the algebraic modeling interface compatible with SIMD abstraction, ExaModels.jl requires users to encode the model equations using a {\tt Generator} data type in Julia.
This data type pairs an instruction (a Julia function) with data (a host or device array) over which the instruction is executed.
This design naturally preserves the NLP model information in the SIMD abstraction described by \eqref{eqn:prob} and facilitates both the evaluation and differentiation of the model on \gls*{gpu} accelerators.
The following code snippet describes the modeling syntax of ExaModels.jl:

\begin{lstlisting}[language=Julia, caption={Modeling in ExaModels.jl}, label={model_exa}]
function luksan_vlcek_model(N; backend = nothing)
  core = ExaCore(backend = backend)
  x = variable(core, N; start = (mod(i, 2) == 1 ? -1.2 : 1.0 for i = 1:N))
  constraint(core, 3x[i+1]^3 + 2 * x[i+2] - 5 + sin(x[i+1] - x[i+2])sin(x[i+1] + x[i+2]) + 4x[i+1] - x[i]exp(x[i] - x[i+1]) - 3 for i = 1:N-2)
  objective(core, 100 * (x[i-1]^2 - x[i])^2 + (x[i-1] - 1)^2 for i = 2:N)
  return ExaModel(core)
end
\end{lstlisting}
{\tt ExaCore} is an object where model cache is stored, {\tt variable} is the function to create variables, {\tt constraint} is the function to create constraints, and {\tt objective} is the function to create objective function.
One can see that
no for loops are used in the model construction, as all such repeated operations are encoded as generator expressions.
Rather, the constraint and objective functions are expressed in the form of {\tt Generator} that reveals the repeated structure.
In the very end, {\tt ExaModel} is created, which provides the callback functions discussed in \Cref{nlpmodels}. One can optionally specify the {\tt backend}, which can be either CPU or \gls*{gpu}.

%%% Local Variables:
%%% mode: LaTeX
%%% TeX-master: "main"
%%% End:

\section{Feature Overview and Implementation of ExaModelsPower.jl}\label{sec:exap}

In this section, we provide an overview of the features and implementation of ExaModelsPower.jl. We first describe how the SIMD abstraction adopted by ExaModels.jl can be used to model power system optimization problems. Next, we present the various models implemented in ExaModelsPower.jl and illustrate how they can be used. Finally, we describe how the users can modify the models via the user callback feature.

\subsection{Design Principles of ExaModelsPower.jl}\label{sec:design}
ExaModelsPower.jl is a modeling library for power system optimization, built on ExaModels.jl. 
It focuses on providing a user-friendly interface for creating \gls*{gpu}-compatible power system optimization models.
The model returns an \texttt{ExaModel} object, which is a subtype of \texttt{AbstractNLPModel} within NLPModels.jl, and thus, can be solved using any solver that is compatible with NLPModels.jl, including MadNLP.jl \cite{shin_accelerating_2024}, Ipopt \cite{wachter_implementation_2006}, KNITRO \cite{pardalos_knitro_2006}, and many others.
Via ExaModels.jl, the model can be evaluated either on CPU or GPU.

\subsection{Model Implementations within ExaModelsPower.jl}
We provide a code snippet that illustrates how ExaModelsPower.jl implements a model using ExaModels.jl syntax.
\Cref{model_breakdown} describes the implementation of static \gls*{ac}\gls*{opf} model in polar coordinates within ExaModelsPower.jl.

\begin{lstlisting}[language=Julia, caption={\gls*{opf} Internal Construction}, label={model_breakdown}]
c = ExaCore() 
  
va = variable(core, length(data.bus))
vm = variable(core, length(data.bus); start = fill!(similar(data.bus, Float64), 1.0), lvar = data.vmin, uvar = data.vmax)
pg = variable(core, length(data.gen); lvar = data.pmin, uvar = data.pmax)
qg = variable(core, length(data.gen); lvar = data.qmin, uvar = data.qmax)
p = variable(core, length(data.arc); lvar = -data.rate_a, uvar = data.rate_a)
q = variable(core, length(data.arc); lvar = -data.rate_a, uvar = data.rate_a)

o = objective(core, gen_cost(g, pg[g.i]) for g in data.gen)

c_ref_angle = constraint(core, c_ref_angle_polar(va[i]) for i in data.ref_buses)
c_to_active_power_flow = constraint(core, c_to_active_power_flow_polar(b, p[b.f_idx], vm[b.f_bus],vm[b.t_bus],va[b.f_bus],va[b.t_bus]) for b in data.branch)
c_to_reactive_power_flow = constraint(core, c_to_reactive_power_flow_polar(b, q[b.f_idx], vm[b.f_bus],vm[b.t_bus],va[b.f_bus],va[b.t_bus]) for b in data.branch)
c_from_active_power_flow = constraint(core, c_from_active_power_flow_polar(b, p[b.t_idx], vm[b.f_bus],vm[b.t_bus],va[b.f_bus],va[b.t_bus]) for b in data.branch)
c_from_reactive_power_flow = constraint(core, c_from_reactive_power_flow_polar(b, q[b.t_idx], vm[b.f_bus],vm[b.t_bus],va[b.f_bus],va[b.t_bus]) for b in data.branch)
c_phase_angle_diff = constraint(core, c_phase_angle_diff_polar(b,va[b.f_bus], va[b.t_bus]) for b in data.branch; lcon = data.angmin, ucon = data.angmax)
c_active_power_balance = constraint(core, c_active_power_balance_demand_polar(b, vm[b.i]) for b in data.bus)
c_reactive_power_balance = constraint(core, c_reactive_power_balance_demand_polar(b, vm[b.i]) for b in data.bus)
c_active_power_balance_arcs = constraint!(core, c_active_power_balance, a.bus => p[a.i] for a in data.arc)
c_reactive_power_balance_arcs = constraint!(core, c_reactive_power_balance, a.bus => q[a.i] for a in data.arc)
c_active_power_balance_gen = constraint!(core, c_active_power_balance, g.bus => -pg[g.i] for g in data.gen)
c_active_power_balance_gen = constraint!(core, c_reactive_power_balance, g.bus => -qg[g.i] for g in data.gen)
c_from_thermal_limit = constraint(core, c_thermal_limit(b,p[b.f_idx],q[b.f_idx]) for b in data.branch; lcon = fill!(similar(data.branch, Float64, length(data.branch)), -Inf))
c_to_thermal_limit = constraint(core, c_thermal_limit(b,p[b.t_idx],q[b.t_idx]) for b in data.branch; lcon = fill!(similar(data.branch, Float64, length(data.branch)), -Inf))

model = ExaModel(core; kwargs...)
\end{lstlisting}

%%% Local Variables:
%%% mode: LaTeX
%%% TeX-master: t
%%% End:

Similarly to \Cref{model_exa}, the model is constructed by first defining the variables, objective, and constraints using the ExaModels.jl standard API.
Note that {\tt constraint!} is used to modify an existing constraint, which is useful when the constraint needs to be modified multiple times, such as in the case of power balance constraints where the contributions from various devices are summed up. For example, the syntax {\tt a.bus=>p[a.i]} adds the term {\tt p[a.i]} to the {\tt a.bus}-th constraint in the original set of constraints.
An \texttt{ExaModel} object is then created using the core object that contains all of the cached model information. 
ExaModelsPower.jl implements all the internal functions required to model the \gls*{opf} in both polar and rectangular coordinates, such as \texttt{gen\_cost}, \texttt{c\_ref\_angle\_polar}, \texttt{c\_to\_active\_power\_flow\_polar}, etc.

\subsection{User Interface and Options}
As of ExaModelsPower.jl v0.1.0, we provide implementations for the static \gls*{opf}, \gls*{mp}\gls*{opf}, and the \gls*{goc3}-variant of \gls*{sc}\gls*{opf}.
For each of these models, ExaModelsPower.jl provides high-level user interface to generate these models. Below, we describe the interfaces for each model with code examples.

\subsubsection{Static \gls*{opf} Models}
The \texttt{opf\_model} function implemented within ExaModelsPower.jl creates the \gls*{ac}\gls*{opf} problem as described in \cite{coffrin_powermodels_2018} using the following one line syntax:
\begin{lstlisting}[language=Julia, caption={User Interface for static OPF models}, label={static}]
using ExaModelsPower, NLPModelsIpopt
model, vars, cons = ac_opf_model("pglib_opf_case118_ieee.m";)
result = ipopt(model)
sol_vm = solution(result, vars.vm)
\end{lstlisting}
The only argument is the instance's name (automatically pulled from pglib-opf \cite{babaeinejadsarookolaee_power_2021} repository) or the path to the data file in the MATPOWER format \cite{zimmerman_matpower_2011}.
Internally, the case data parsing is handled by ExaPowerIO.jl, which provides highly efficient data parsing capabilities.
By default, the model is constructed in polar coordinates, but this can be changed by specifying the {\tt form} keyword argument.
The model function returns the model object, as well as the references to variables and constraints grouped in named tuples for easy access.
Here, Ipopt \cite{wachter_implementation_2006} is used as the solver.
The entire solution is returned by the solver (the standard convention within NLPModels.jl ecosystem), and the solution for specific variable can be queried using the \texttt{solution} function.
In \Cref{static}, \texttt{sol\_vm} contains the optimal voltage magnitude at each bus.

The example in \Cref{static_gpu} illustrates how the model can be formulated in rectangular form and solved on NVIDIA \gls*{gpu}:
\begin{lstlisting}[language=Julia, caption={User Interface for static OPF models}, label={static_gpu}]
using ExaModelsPower, MadNLPGPU, CUDA
model, vars, cons = ac_opf_model("pglib_opf_case118_ieee.m"; backend = CUDABackend(), form = :rect)
result = madnlp(model; tol=1e-6)
\end{lstlisting}
In this example, the user has also specified the coordinate system (rectangular) using the \texttt{form} keyword argument and
the hardware backend using the \texttt{backend} keyword argument.
The problem is then solved using MadNLP.jl on GPU.
Using multiple dispatch, the \texttt{madnlp} function automatically identifies the relvant hardware backend and dispatches the computation within MadNLP.jl on the GPU \cite{shin_accelerating_2024}. 

\subsubsection{\Gls*{mp}\gls*{opf} Models}
The \texttt{mpopf\_model} function implemented within ExaModelsPower.jl creates the multi-period \gls*{ac}\gls*{opf} problem as described in \cite{shin_scalable_2024}.
This model formulates the \gls*{ac}\gls*{opf} over multiple time periods, linking the time periods through ramping constraints on each generator.
Depending on how we specify the temporal variation in load, ExaModelsPower.jl provides two different user interfaces for creating the \gls*{mp}\gls*{opf} model.

First, the following code snippet illustrates an interface where the user specifies a representative demand curve that scales the static demand at each bus across time points.
\begin{lstlisting}[language=Julia, caption={User Interface for \gls*{mp}\gls*{opf} models}, label={mp_curve}]
curve = [1, .9, .8, .85] # Demand scaling curve
model, vars, cons = mpopf_model("pglib_opf_case118_ieee.m", curve; corrective_action_ratio = 0.2)
\end{lstlisting}
Here,
\texttt{mpopf\_model} takes two main arguments: the name of the data file and a vector that specifies the representative demand curve.
The temporal demand is created by multiplying the ratio at each point in the input demand vector by the static demand given in the data file. Demand at each bus is scaled by the same ratio.
The keyword argument \texttt{corrective\_action\_ratio} specifies the permissible difference in active power output at each generator across two consecutive time points \cite[(1k)]{shin_scalable_2024}.

Alternatively, we provide an alternative API to directly provide active/reactive power load at each bus and each time point in the form of time series:
\begin{lstlisting}[language=Julia, caption={Alternative user interface for \gls*{mp}\gls*{opf} models}, label={mp_granular}]
model, vars, cons = mpopf_model(
    "pglib_opf_case3_lmbd.m", "case3.Pd", "case3.Qd")
\end{lstlisting}
Here, \texttt{case3.Pd} and \texttt{case3.Qd} are paths to files that specify the power loads.

\subsubsection{\Gls*{mp}\gls*{opf} Models with Storage}
ExaModelsPower.jl provides further options by extending the \gls*{mp}\gls*{opf} to allow for the inclusion of storage as formulated by \cite{geth_flexible_2020} which includes the modeling of the state of charge of storage units, as well as charging and discharging efficiencies. The storage model is created automatically if the input data file includes storage parameters. 

A complication introduced by this formulation is the complementarity constraint: $P_{c,k}^c \cdot P_{c,k}^d = 0$, where $P_{c,k}^c$ and $P_{c,k}^d$ are the charging and discharging rate at device $c$ and time $k$, respectively.
Directly enforcing complementarity constraints as nonlinear equality constraints can lead to numerical instability \cite{nurkanovic2023solving}.
One way to handle such complementarity constraint is to simply relax them and check the constraint violation after the solution is obtained.
The relaxed formulation can be created using the following user interface:
\begin{lstlisting}[language=Julia, caption={User interface for \gls*{mp}\gls*{opf} models with storage}, label={storage}]
  # Include complementarity constraint
  model, vars, cons = mpopf_model("pglib_opf_case5_pjm_mod.m", curve; storage_complementarity_constraint = true)
\end{lstlisting}
Here, \texttt{storage\_complementarity\_constraint} is set to true to relax the complementarity constraint.

\subsubsection{\Gls*{goc3} Models}

ExaModelsPower.jl further provides modeling of the security constrained problem as presented in the \gls*{goc3}.
\gls*{goc3} formulation include lots of additional components beyond the standard \gls*{opf} formulation as in \Cref{model_breakdown}, including contingency constraints, flexible loads, reserve requirements, unit commitment decisions, and many others \cite{holzer_grid_2024}.
The \gls*{goc3} formulation includes a total of 163 unique constraints, including unit commitment decisions, ramping limits, transformers, and reserve requirements.
We model the continuous part of the \gls*{goc3} formulation, where the unit commitment decisions are specified already.

\Cref{goc3} describes the user interface for creating the \gls*{goc3} model using the \texttt{goc3\_model} function.
\begin{lstlisting}[language=Julia, caption={GOC3 modeling}, label= {goc3}]
# Construct using UC solution from benchmark solver
model, cons, vars, lengths, sc_data_array = goc3_model("data/C3E4N00073D1_scenario_303.json", "data/C3E4N00073D1_scenario_303_solution.json")
\end{lstlisting}
The case data provided in JSON format, as specified by the ARPA-E Grid Optimization Competition \cite{stephen_elbert_arpa-e_2024}, and the unit commitment decisions in JSON format are used as the input. 

\subsection{User Extensions}
ExaModelsPower.jl also supports user extensions to existing models, implemented by a user-defined callback function during the model construction.
This callback function is called after the base model is constructed, allowing the user to perform arbitrary modifications to the model.
The callback function takes three arguments: the core object that stores the model cache, and the named tuples of variables and constraints that store references to the variables and constraints created by ExaModelsPower.jl.
The user can add new variables, objectives, and constraints to the model using the ExaModels.jl interface.
\Cref{extend} provides an example of extending the \gls*{mp}\gls*{opf} with electrolyzers connected to the network.

\begin{lstlisting}[language=Julia, caption={Modeling User Extension}, label={extend}]
curve = [1, .9, .95]
# Create vector of NamedTuples elec\_data w/ device data
untimed_elec_data = [(i = 1, bus = 1, cost = -5000), (i = 2, bus = 2, cost = -2000)]
Ntime = 3; Nbus = 2
elec_data = [(;b..., t = t) for b in untimed_elec_data, t in 1:Ntime]
elec_min = zeros(size(elec_data)); elec_max = fill(50, size(elec_data)); elec_scale = Float64(10)

# User-defined model modifications go here
function add_electrolyzers(core, vars, cons)
    # Add new variable to core
    p_elec = variable(core, size(elec_data, 1), size(elec_data, 2); lvar = elec_min, uvar = elec_max)
    
    # Objectives are additive. Add secondary objective
    o2 = objective(core, e.cost*p_elec[e.i, e.t] for e in elec_data)
    
    # Add electrolyzer load to power balance for each bus
    c_elec_power_balance = constraint!(core, cons.c_active_power_balance, e.bus + Nbus*(e.t-1) => p_elec[e.i, e.t] for e in elec_data)
    
    # Ramping limit over time
    c_elec_ramp = constraint(core, p_elec[e.i, e.t] - p_elec[e.i, e.t - 1] for e in elec_data[:, 2:Ntime]; lcon = fill(-elec_scale, size(elec_data[:, 2:Ntime])), ucon = fill(elec_scale, size(elec_data[:, 2:Ntime])))
    
    # Set initial electrolyzer power to 0
    c_elec_ramp_init = constraint(core, p_elec[e.i, e.t] for e in elec_data[:, 1];)

    # Return new variables and constraints to be tracked
    return ((p_elec=p_elec,), (c_elec_ramp=c_elec_ramp, c_elec_ramp_init=c_elec_ramp_init))
end
# Generate model
model, vars, cons = mpopf_model("pglib_opf_case3_lmbd.m", curve; user_callback = add_electrolyzers)
\end{lstlisting}

In this example, one can see that the new set of variables representing the power consumption of the electrolyzers are added to the model, along with an additional objective that minimizes the cost of hydrogen production, a modification of the active power balance constraint to include the electrolyzer load, and a new ramping constraint on the electrolyzer power consumption.
Our current model implementations \texttt{opf\_model} and \texttt{mpopf\_model} both accept a keyword argument \texttt{user\_callback}, which takes a user-defined function as its value.

%%% Local Variables:
%%% mode: LaTeX
%%% TeX-master: "main"
%%% End:

\section{Benchmarking Results}\label{sec:num}

This section demonstrates the core modeling functionality of ExaModelsPower.jl and provides an overview of the current capabilities of \gls*{gpu}-accelerated optimization solvers for nonlinear power system optimization problems. The results presented here can be reproduced with the code available at \url{https://github.com/mit-shin-group/exa-power-pscc-2026}.

\subsection{Benchmark Setup}
\subsubsection{Hardware Configurations}
Benchmarks are conducted on a workstation with two Intel Xeon Gold 6130 CPUs, two NVIDIA Quadro GV100 \glspl*{gpu}, and 1 TB of RAM.

\subsubsection{ExaModelsPower.jl Configurations}
ExaModelsPower.jl v0.1.0 is used for all benchmarks. As described in \Cref{sec:exap}, ExaModelsPower.jl can evaluate model functions and derivatives on both CPU and \gls*{gpu} backends. 
We use the CPU backend for benchmarking CPU solvers and 
the CUDA backend for benchmarking \gls*{gpu} solvers. For CPU configurations, ExaModels.jl runs in single-threaded mode, as parallelism at the model- and derivative-evaluation level does not provide a significant performance improvement on CPU.

\subsubsection{Solver Configurations}
The following solvers are used in the benchmarks:
\begin{itemize}[leftmargin=*,itemsep=0pt,topsep=0pt,partopsep=0pt,parsep=0pt]
\item \textbf{Ipopt (CPU)---Ipopt v3.14.19 with ma27 on CPU} \cite{wachter_implementation_2006}: Ipopt is a widely used open-source CPU-based interior-point solver. When it is configured with HSL ma27 linear solver, the entire solver runs on a single thread. For static \gls*{ac} \gls*{opf}, Ipopt with ma27 is among the fastest open-source solvers \cite{tasseff_exploring_2019}. Ipopt can reliably find the local optimal solution of static \gls*{ac} \gls*{opf} problems with $10^{-8}$ tolerance (based on relative error to KKT conditions). 
\item \textbf{MadNLP (LK, CPU)---MadNLP.jl v0.8.10 with LiftedKKT and ma86 on CPU} \cite{shin_accelerating_2024}: MadNLP.jl is an interior-point solver that can run on both CPU and \gls*{gpu}, and its algorithm is based on the filter line-search method used by Ipopt \cite{wachter_implementation_2006}. MadNLP.jl supports various KKT systems beyond the augmented system used in Ipopt, including the LiftedKKT \cite{pacaud2024condensed} and HybridKKT \cite{regev_hykkt_2023}. LiftedKKT is particularly well-suited for linear solvers that exploit multi-thread parallelism at the elimination-tree level, such as ma86, because the matrix treated by direct solver is always made positive definite, thus eliminating the need for numerical pivoting. More details on the algorithmic differences between LiftedKKT, HybridKKT and the augmented system can be found in \cite{pacaud2024condensed}. 
In this configuration, MadNLP.jl runs on a single thread, but the linear solver ma86 can utilize multiple threads. LiftedKKT slightly relaxes the equality constraints by the solver tolerance; thus, when loose tolerance is used, the final solution may have a non-negligible constraint violation.
\item \textbf{MadNLP (LK, GPU)---MadNLP.jl v0.8.10 with LiftedKKT and cuDSS on GPU} \cite{shin_accelerating_2024}: This configuration runs on an NVIDIA \gls*{gpu} and solves sparse linear systems using cuDSS. By default, MadNLP.jl uses the LiftedKKT formulation when the backend is set to \gls*{gpu}. MadNLP.jl shares a common code base for CPU and \gls*{gpu} versions. However, the low-level kernels are specialized for the hardware backend via Julia's multiple dispatch, ensuring that the performance-critical parts are optimized for \gls*{gpu} execution. Thus, this configuration of solver will run the same algorithm as the MadNLP+LiftedKKT+ma86 CPU configuration, but everything is executed on the \gls*{gpu} instead.
\item \textbf{MadNLP (HK, GPU)---MadNLP.jl v0.8.10 with HybridKKT and cuDSS on GPU} \cite{shin_accelerating_2024}: HybridKKT \cite{regev_hykkt_2023} is used in this configuration instead of LiftedKKT. Most algorithmic details are the same as in the LiftedKKT \gls*{gpu} configuration. HybridKKT does not apply the equality relaxation, and thus, is able to achieve better constraint satisfaction than LiftedKKT for the same convergence tolerance; however, HybridKKT internally relies on an iterative solver to solve the Schur complement system, which may lead to slower performance than LiftedKKT in some cases.
\item \textbf{MadNCL (GPU)---MadNCL.jl v0.1.0 with cuDSS on GPU} \cite{montoison_madncl_2025}: MadNCL.jl is a new augmented-Lagrangian–based solver that runs on both CPU and \gls*{gpu}. Like MadNLP.jl, MadNCL.jl shares a common code base for CPU and \gls*{gpu} versions, but low-level kernels are compiled differently for each hardware backend. 
\end{itemize}

\subsubsection{Shifted Geometric Mean}
We use wall-clock time when reporting times. For summaries across multiple instances, we employ the shifted geometric mean (SGM). Let $t_i$ denote the wall-clock time for instance $i$, with $n$ instances in total, and $\Delta$ the shift parameter (set to 10 seconds in this work to reduce the influence of very small solve times). The SGM is defined as $t_{\mathrm{sgm}} = \left(\prod_{i=1}^{n} (t_i + \Delta)\right)^{1/n} - \Delta$.
We refer to SGM10 as the shifted geometric mean with $\Delta = 10$ seconds.
If an instance is unsolved or timed out, its solve time is set to the maximum time limit.

\subsubsection{Constraint Violation}
The constraint violations are computed as $\max\left(\left\|g^\flat - g(x^\star)\right\|_\infty,\left\|g(x^\star) - g^\sharp\right\|_\infty\right)$, where $g^\flat$ and $g^\sharp$ are the constraints lower and upper bounds, and $x^\star$ is the final solution returned by the solver.
We apply the SGM10 to compare constraint violations across instances.
If an instance is unsolved, its constraint violation was not included in the calculation of the SGM10 constraint violation.

\subsection{Static \gls*{opf} Formulation}\label{sec:num-static}

Static and multi-period \gls*{ac}\gls*{opf} cases were collected from the PGLib-\gls*{opf} library \cite{babaeinejadsarookolaee_power_2021} (version 23.07). We report all base cases in this library, excluding the \texttt{api} (heavily loaded test cases) and \texttt{sad} (small phase-angle-difference cases) instances.

\begin{figure}[t]
    \centering
    \includegraphics[width=1\linewidth]{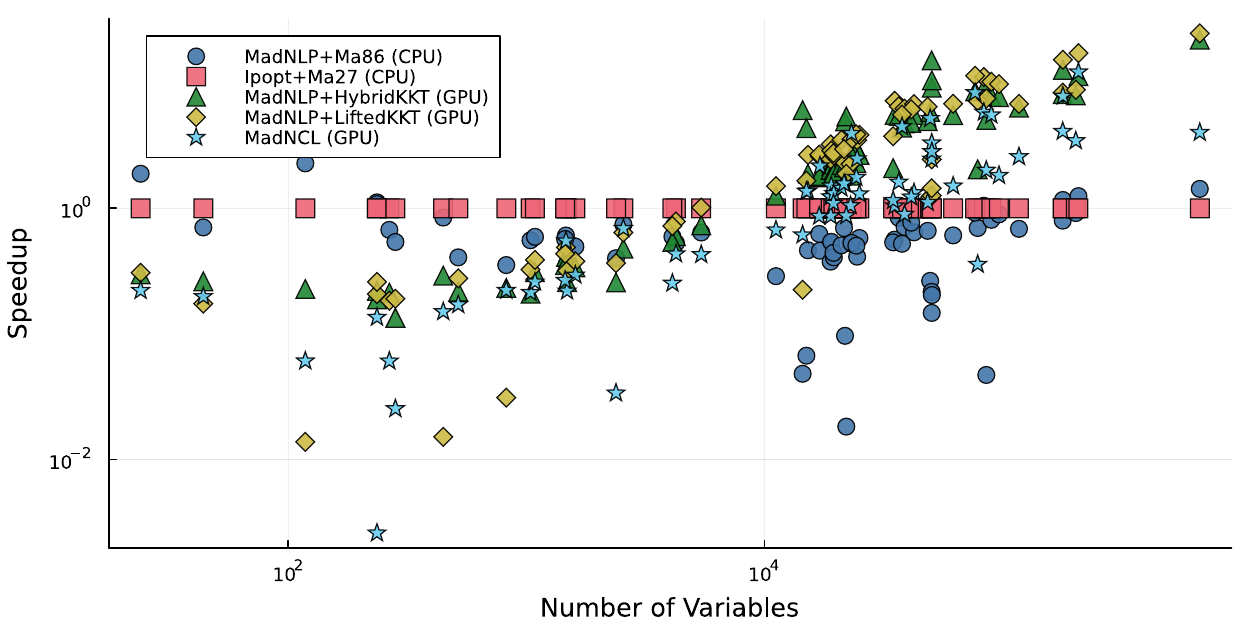}
    \caption{Speedup of different configurations of solvers for static OPFs, compared to Ipopt+ma27 running on CPU.}
    \label{Static speedup}
\end{figure}

\Cref{Summary of Static OPF Benchmarking} presents the solver performances for the static \gls*{opf} modeled in polar and rectangular coordinates. For both setups, we use a 900-second wall-clock time limit. \Cref{Static speedup} shows the speedup achieved by different solvers as problem size grows, using the case of Ipopt configured with ma27 running on CPU as a baseline. 

Key observations are as follows:
(1) When the SGM10 time is compared across all problem sizes, MadNLP+LiftedKKT yields the best performance for moderate tolerance, but MadNCL performs better when a polar formulation is used and a tight tolerance of $10^{-8}$ is required.
(2) When we focus on large problems only, MadNLP on GPU is the fastest for both polar and rectangular formulations, and for both tolerances.
(3) For small problems, CPU solvers are consistently faster than \gls*{gpu} solvers. This result is expected, as the advantages of GPU computation are lower on small cases which allow for a smaller degree of parallelization.
(4) All three \gls*{gpu} solvers experience more frequent failures under tighter tolerances, while CPU solvers show only one additional failure.
(5) Solve times are almost always shorter when the problem is modeled in polar coordinates, except for small problems. 
 
\setlength{\textfloatsep}{5pt}

\setlength{\tabcolsep}{5pt}
\begin{table*}[t] 
\centering
\renewcommand{\arraystretch}{.95}
\caption{Summary of Static OPF Benchmarking}
\label{Summary of Static OPF Benchmarking}
\begin{tabular}{|c|c|c|rrr|rrr|rrr|rrr|}
\hline
 & & & \multicolumn{3}{c|}{\textbf{Small (23)}} & \multicolumn{3}{c|}{\textbf{Medium (31)}} & \multicolumn{3}{c|}{\textbf{Large (12)}} & \multicolumn{3}{c|}{\textbf{Total (66)}} \\
  & \textbf{tol} & \textbf{Solver} & \multicolumn{3}{c|}{\textbf{nvar$<$15k}} & \multicolumn{3}{c|}{\textbf{15k$<$nvar$<$75k}} & \multicolumn{3}{c|}{\textbf{nvar$>$75k}} &  \multicolumn{3}{c|}{}\\
 & & & \textbf{Count} & \textbf{Time} & \textbf{Cvio} & \textbf{Count} & \textbf{Time} & \textbf{Cvio} & \textbf{Count} & \textbf{Time} & \textbf{Cvio} & \textbf{Count} & \textbf{Time} & \textbf{Cvio}\\
\hline
\multirow{10}{*}{\rotatebox{90}{\textbf{Polar}}}
 & \multirow{5}{*}{\textbf{$10^{-4}$}} & \textbf{MadNLP (LK, GPU)} & \textbf{23} & 0.82 & 1.38e-3 & \textbf{31} & \textbf{1.47} & 1.47e-3 & \textbf{12} & \textbf{3.42} & 2.18e-3 & \textbf{66} & \textbf{1.57} & 1.57e-3 \\
 & & \textbf{MadNLP (HK, GPU)} & 22 & 2.49 & \textbf{8.76e-7} & 23 & 24.0 & \textbf{2.61e-7} & \textbf{12} & 4.45 & \textbf{1.21e-7} & 57 & 10.53 & \textbf{4.69e-7} \\
 & & \textbf{MadNCL (GPU)} & 21 & 5.64 & 8.70e-5 & \textbf{31} & 2.32 & 5.38e-3 & \textbf{12} & 10.73 & 8.62e-4 & 64 & 4.72 & 2.80e-3 \\
 & & \textbf{Ipopt (CPU)} & \textbf{23} & \textbf{0.25} & 1.88e-6 & \textbf{31} & 3.75 & 7.57e-6 & \textbf{12} & 33.26 & 5.37e-6 & \textbf{66} & 5.29 & 5.18e-6 \\
 & & \textbf{MadNLP (LK, CPU)} & 22 & 2.46 & 1.93e-6 & \textbf{31} & 10.72 & 1.05e-6 & 11 & 50.16 & 8.64e-7 & 64 & 11.06 & 1.32e-6 \\
\cline{2-15}
& \multirow{5}{*}{\textbf{$10^{-8}$}} & \textbf{MadNLP (LK, GPU)} & 22 & 2.86 & 1.36e-7 & \textbf{31} & 4.12 & 1.45e-7 & 9 & 35.25 & 1.17e-7 & 62 & 6.89 & \textbf{1.38e-7} \\
 & & \textbf{MadNLP (HK, GPU)} & 22 & 2.53 & \textbf{1.35e-7} & 23 & 25.22 & \textbf{1.28e-7} & 10 & \textbf{22.22} & 2.41e-7 & 55 & 14.17 & 1.51e-7 \\
 & & \textbf{MadNCL (GPU)} & 19 & 13.34 & 1.55e-7 & \textbf{31} & \textbf{3.0} & 2.06e-7 & 7 & 87.63 & \textbf{1.16e-7} & 57 & 13.0 & 1.78e-7 \\
 & & \textbf{Ipopt (CPU)} & \textbf{23} & \textbf{0.27} & 1.35e-7 & \textbf{31} & 4.26 & 1.44e-7 & \textbf{12} & 37.23 & 2.18e-7 & \textbf{66} & \textbf{5.81} & 1.55e-7 \\
 & & \textbf{MadNLP (LK, CPU)} & 22 & 2.66 & 1.37e-7 & \textbf{31} & 15.82 & 1.46e-7 & 10 & 88.27 & 1.20e-7 & 63 & 15.68 & 1.38e-7 \\
\hline
 \multirow{10}{*}{\rotatebox{90}{\textbf{Rectangular}}} & \multirow{5}{*}{\textbf{$10^{-4}$}} & \textbf{MadNLP (LK, GPU)} & \textbf{23} & 0.4 & 1.39e-3 & \textbf{30} & \textbf{3.09} & 1.46e-3 & \textbf{10} & \textbf{18.47} & 2.36e-3 & \textbf{63} & \textbf{3.91} & 1.58e-3 \\
 & & \textbf{MadNLP (HK, GPU)} & 22 & 2.55 & \textbf{9.89e-7} & 22 & 29.51 & 1.39e-6 & 9 & 31.64 & \textbf{1.43e-7} & 53 & 16.75 & \textbf{1.01e-6} \\
 & & \textbf{MadNCL (GPU)} & 21 & 5.43 & 1.08e-4 & 29 & 6.86 & 3.92e-3 & 5 & 163.18 & 1.89e-3 & 55 & 14.97 & 2.28e-3 \\
 & & \textbf{Ipopt (CPU)} & \textbf{23} & \textbf{0.17} & 1.98e-6 & \textbf{30} & 6.29 & 8.83e-6 & 9 & 78.23 & 4.95e-6 & 62 & 8.79 & 5.73e-6 \\
 & & \textbf{MadNLP (LK, CPU)} & \textbf{23} & 0.86 & 1.64e-6 & \textbf{30} & 12.1 & \textbf{1.26e-6} & 8 & 115.17 & 1.08e-6 & 61 & 13.65 & 1.38e-6 \\
\cline{2-15}
& \multirow{5}{*}{\textbf{$10^{-8}$}} & \textbf{MadNLP (LK, GPU)} & 22 & 2.73 & 1.37e-7 & 29 & \textbf{6.36} & 1.40e-7 & 8 & \textbf{50.44} & 2.26e-7 & 59 & \textbf{9.01} & 1.51e-7 \\
 & & \textbf{MadNLP (HK, GPU)} & 22 & 2.62 & 1.35e-7 & 22 & 31.62 & \textbf{1.28e-7} & 7 & 86.01 & 2.85e-7 & 51 & 21.97 & 1.53e-7 \\
 & & \textbf{MadNCL (GPU)} & 21 & 5.64 & \textbf{1.31e-7} & 29 & 7.56 & 1.60e-7 & 4 & 236.94 & \textbf{1.06e-7} & 54 & 17.27 & \textbf{1.45e-7} \\
 & & \textbf{Ipopt (CPU)} & \textbf{23} & \textbf{0.2} & 1.35e-7 & \textbf{30} & 6.76 & 1.45e-7 & \textbf{9} & 83.34 & 2.47e-7 & \textbf{62} & 9.26 & 1.56e-7 \\
 & & \textbf{MadNLP (LK, CPU)} & \textbf{23} & 1.21 & 1.38e-7 & \textbf{30} & 14.14 & 1.45e-7 & 8 & 134.84 & 2.26e-7 & 61 & 15.59 & 1.53e-7 \\
\hline
\end{tabular}
\caption*{\footnotesize Cvio = maximum constraint violation, nvar = Number of Variables, LK = LiftedKKT, HK = HybridKKT.}
\end{table*}

\subsection{\Gls*{mp} \gls*{opf} Formulation (with or without storage)}\label{sec:num-mp}
We have used polar coordinates for all \gls*{mp}\gls*{opf} cases, as they consistently yield better performance in the static \gls*{opf} benchmarks.
The time-dependent load profiles were generated by equally varying the demand at each bus by the same curve, based on typical power demand from \cite{grigg_ieee_1999} for a typical summer weekday, with the \texttt{corrective\_action\_ratio} (defined in \Cref{mp_curve}) set to $0.25$.
It is important to note that these models are not guaranteed to be feasible, and checking the feasibility in a global sense is computationally intractable.
As such, we exclude the cases where all solvers fail to find a feasible solution within the time limit from the summary figures and statistics.

The \gls*{mp}\gls*{opf} with storage cases are generated by building on the multi-period cases described above.
We place storage devices at the buses with largest 10\% active power load within the network.
The energy rating of each storage device is set as 1.5 times the power demand of the given bus, loosely based on the ratio of 1.27 currently observed in CAISO \cite{department_of_market_monitoring_2024_2025, california_iso_gross_2023}.
The ratio of energy rating to charge and discharge rating is specified to match the test parameters used in \cite{geth_flexible_2020}.
This setup is chosen to demonstrate a basic test case that implements storage modeling, and is not intended to represent realistic conditions.
Because no negative generation costs are used, no complementarity constraints are enforced, and cases for which every solver failed were also dropped from the summary figures for \gls*{mp} with storage benchmark results.
For both setups, we used a 900-second wall-clock time limit.

\begin{figure}[h]
  \centering
  \includegraphics[width=1\linewidth]{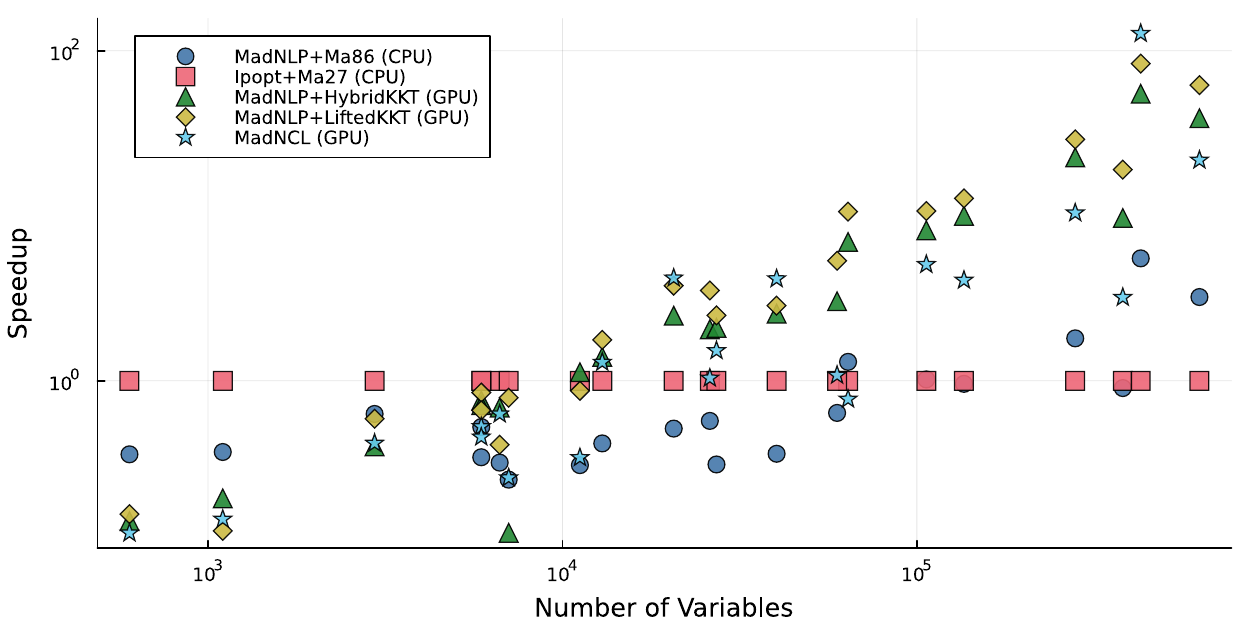}
    \caption{Speedup of different configurations of solvers for \gls*{mp} OPFs, compared to Ipopt+ma27 running on CPU.}
    \label{Mp speedup}
\end{figure}

\begin{figure}[t]
  \centering
  \includegraphics[width=1\linewidth]{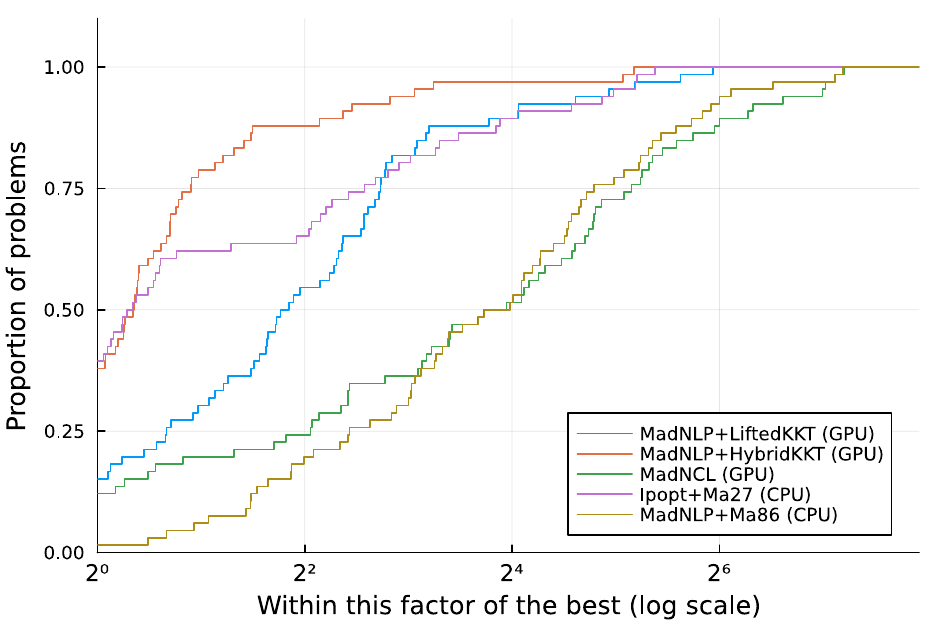}
    \caption{Proportion of \gls*{mp} OPF Cases Solved Within Factor of Fastest Solver}
    \label{Mp compare}
\end{figure}

\Cref{Summary of MPOPF Benchmarking} presents the benchmark results for the MPOPF cases with and without storage.
\Cref{Mp speedup} depicts the speedup achieved by different solvers as problem size increases for the MPOPF without storage. \Cref{Mp compare} presents the speed of various solvers as a factor of the fastest solver across cases.
The results for the \gls*{mp}\gls*{opf} benchmarking are summarized as so:
(1) Due to the time limit, \gls*{gpu} solvers have much higher success rate than CPU solvers, particularly for large instances. 
Furthermore, within the given time limit, large-sized cases with storage are only solvable using \gls*{gpu} solvers.
(2) Adding storage decreases the amount of cases that any solver can solve, especially for large cases at high tolerance. A large portion of these failures arise from max wall time violations.
(3) \gls*{gpu} solvers achieve nearly 2 orders of magnitude speedup for the largest instances that are solvable on CPU in the given time limit.
(4) \gls*{gpu} solvers generate the fastest solution for approximately two-thirds of cases.
(5) At both tolerances, multi-thread CPU parallelism via ma86 provides substantial speed-ups over the single-threaded solver Ipopt with ma27.
(6) LiftedKKT and MadNCL.jl exhibit larger constraint violations compared to the other solvers.

\setlength{\textfloatsep}{5pt}

\setlength{\textfloatsep}{5pt}

\setlength{\tabcolsep}{5pt}
\begin{table*}[t] 
\centering
\renewcommand{\arraystretch}{0.95}
\caption{Summary of MPOPF Benchmarking}
\label{Summary of MPOPF Benchmarking}
\renewcommand{\arraystretch}{0.95}
\begin{tabular}{|c|c|c|rrr|rrr|rrr|rrr|}
\hline
  & \multirow{3}{*}{\textbf{tol}} & \multirow{3}{*}{\textbf{Solver}}& \multicolumn{3}{c|}{\textbf{Small (16)}} & \multicolumn{3}{c|}{\textbf{Medium (16)}} & \multicolumn{3}{c|}{\textbf{Large (34)}} & \multicolumn{3}{c|}{\textbf{Total (66)}} \\
  & & & \multicolumn{3}{c|}{\textbf{nvar$<$50k}} & \multicolumn{3}{c|}{\textbf{50k$<$nvar$<$500k}} & \multicolumn{3}{c|}{\textbf{nvar$>$500k}} &  \multicolumn{3}{c|}{}\\
 & & & \textbf{Count} & \textbf{Time} & \textbf{Cvio} & \textbf{Count} & \textbf{Time} & \textbf{Cvio} & \textbf{Count} & \textbf{Time} & \textbf{Cvio} & \textbf{Count} & \textbf{Time} & \textbf{Cvio}\\
\hline
\multirow{10}{*}{\rotatebox{90}{\textbf{No Storage}}} & \multirow{5}{*}{\textbf{$10^{-4}$}} & \textbf{MadNLP (LK, GPU)} & \textbf{13} & \textbf{0.4} & 5.28e-4 & \textbf{9} & \textbf{7.09} & 2.59e-3 & \textbf{18} & \textbf{59.41} & 1.95e-3 & \textbf{40} & \textbf{17.33} & 1.63e-3 \\
 & & \textbf{MadNLP (HK, GPU)} & \textbf{13} & 0.62 & \textbf{8.19e-7} & 8 & 18.09 & 7.34e-7 & 12 & 215.9 & \textbf{7.57e-7} & 33 & 42.32 & \textbf{7.76e-7} \\
 & & \textbf{MadNCL (GPU)} & \textbf{13} & 0.63 & 2.36e-5 & 8 & 51.22 & 1.14e-2 & 14 & 225.28 & 3.89e-3 & 35 & 53.51 & 4.16e-3 \\
 & & \textbf{Ipopt  (CPU)} & \textbf{13} & 0.66 & 2.49e-6 & 7 & 132.77 & 8.41e-6 & 1 & 894.61 & 1.03e-6 & 21 & 131.0 & 4.40e-6 \\
 & & \textbf{MadNLP (LK, CPU)} & \textbf{13} & 1.53 & 1.07e-6 & 8 & 86.38 & \textbf{4.93e-7} & 2 & 832.88 & 1.40e-5 & 23 & 118.27 & 1.99e-6 \\
\cline{2-15}
& \multirow{5}{*}{\textbf{$10^{-8}$}} & \textbf{MadNLP (LK, GPU)} & \textbf{13} & 0.95 & 5.28e-8 & \textbf{8} & \textbf{13.64} & 2.70e-7 & \textbf{12} & \textbf{239.19} & \textbf{1.56e-7} & \textbf{33} & \textbf{35.57} & 1.43e-7 \\
&  & \textbf{MadNLP (HK, GPU)} & \textbf{13} & \textbf{0.52} & 4.99e-8 & 6 & 40.03 & \textbf{1.83e-7} & 8 & 259.11 & 1.86e-7 & 27 & 44.95 & \textbf{1.20e-7} \\
 & & \textbf{MadNCL (GPU)} & 12 & 5.0 & 1.74e-7 & 5 & 89.39 & 4.15e-7 & 5 & 476.13 & 5.53e-7 & 22 & 82.91 & 3.15e-7 \\
 & & \textbf{Ipopt  (CPU)} & \textbf{13} & 0.89 & \textbf{4.95e-8} & 6 & 126.28 & 3.08e-7 & 0 & - & - & 19 & 103.93 & 1.31e-7 \\
 & & \textbf{MadNLP (LK, CPU)} & \textbf{13} & 3.39 & 5.28e-8 & 7 & 96.23 & 2.94e-7 & 1 & 833.42 & 4.19e-7 & 21 & 102.73 & 1.51e-7 \\
\hline
\multirow{10}{*}{\rotatebox{90}{\textbf{Storage}}} & \multirow{5}{*}{\textbf{$10^{-4}$}} & \textbf{MadNLP (LK, GPU)} & \textbf{16} & 3.13 & 5.70e-4 & 8 & \textbf{100.59} & 1.15e-3 & \textbf{15} & \textbf{305.46} & 8.94e-4 & \textbf{39} & \textbf{60.65} & 8.14e-4 \\
 & & \textbf{MadNLP (HK, GPU)} & 10 & 49.8 & 2.35e-6 & 0 & - & - & 0 & - & - & 10 & 304.52 & 2.35e-6 \\
 & & \textbf{MadNCL (GPU)} & \textbf{16} & \textbf{2.35} & 1.26e-4 & \textbf{9} & 137.53 & 8.66e-3 & 8 & 714.84 & \textbf{5.88e-4} & 33 & 90.32 & 2.56e-3 \\
 & & \textbf{Ipopt  (CPU)} & 15 & 12.14 & 1.83e-6 & 4 & 627.02 & \textbf{9.81e-8} & 0 & - & - & 19 & 185.63 & 1.46e-6 \\
 & & \textbf{MadNLP (LK, CPU)} & 15 & 14.64 & \textbf{8.27e-7} & 5 & 387.67 & 1.78e-7 & 0 & - & - & 20 & 171.85 & \textbf{6.65e-7} \\
\cline{2-15}
& \multirow{5}{*}{\textbf{$10^{-8}$}} & \textbf{MadNLP (LK, GPU)} & 14 & 11.87 & 5.51e-8 & \textbf{6} & \textbf{119.75} & 1.53e-7 & \textbf{1} & 861.23 & 8.84e-8 & \textbf{21} & \textbf{38.34} & 8.47e-8 \\
 & & \textbf{MadNLP (HK, GPU)} & 12 & 28.26 & \textbf{2.72e-8} & 0 & - & - & 0 & - & - & 12 & 109.73 & \textbf{2.72e-8} \\
 & & \textbf{MadNCL (GPU)} & 14 & 10.57 & 5.44e-8 & 4 & 297.69 & 1.14e-7 & \textbf{1} & \textbf{821.55} & \textbf{6.39e-8} & 19 & 48.98 & 6.73e-8 \\
 & & \textbf{Ipopt  (CPU)} & 15 & 16.68 & 5.14e-8 & 1 & 642.34 & 6.23e-8 & 0 & - & - & 16 & 76.59 & 5.21e-8 \\
 & & \textbf{MadNLP (LK, CPU)} & \textbf{16} & \textbf{10.1} & 5.66e-8 & 4 & 446.25 & 8.86e-8 & 0 & - & - & 20 & 55.37 & 6.30e-8 \\
\hline
\end{tabular}
\caption*{\footnotesize Cvio = maximum constraint violation, nvar = Number of Variables, LK = LiftedKKT, HK = HybridKKT.}
\end{table*}

\subsection{\gls*{goc3} Formulation}\label{sec:num-goc}

The benchmarking results for the \gls*{goc3} formulation are presented in this section.
The original problem formulation aims to solve unit commitment and \gls*{ac}\gls*{opf} simultaneously over a 24 hour horizon, with $N-1$ security constraints for each generator and branch contingency.
We have utilized the UC solutions (binary solutions) generated from \cite{parker_managing_2024}.

\Cref{Summary of GOC3 Benchmarking} presents limited benchmarking results for the \gls*{goc3} security constrained formulation.
While the  time limit for the competition was set to 120 minutes, the limits for this benchmark are set to 500 minutes and 10,000 iterations. The results for the \gls*{goc3} benchmark are summarized as follows: (1) The LiftedKKT solver is the only solver succeeding on all 4 cases, although with a lower constraint violation than the other solvers. (2) While only the smallest case is solvable by other solvers, both \gls*{gpu} KKT methods are faster than the CPU solvers. (3) Without simplifications or decomposition methods, the full \gls*{goc3} formulation cannot be solved at the desired time or scale. Memory errors cause all solvers to fail for 617 bus cases (the competition had cases up to 24,000 buses), nor do any solvers succeed at a tolerance of $10^{-6}$. 

\begin{table}[t] 
\centering
\renewcommand{\arraystretch}{0.9}
\caption{Summary of GOC3 Benchmarking}
\label{Summary of GOC3 Benchmarking}
\renewcommand{\arraystretch}{0.9} 
\begin{tabular}{|c|c|ccc|}
\hline
\multirow{2}{*}{\textbf{tol}} & \multirow{2}{*}{\textbf{Solver}} & \multicolumn{3}{c|}{\textbf{Total (4)}} \\
 & & \textbf{Count} & \textbf{Time} & \textbf{Cvio}\\
\hline
\multirow{5}{*}{\textbf{$10^{-4}$}} & \textbf{MadNLP (LK, GPU)} & \textbf{4} & \textbf{5294.44} & 7.167e-4 \\
 & \textbf{MadNLP (HK, GPU)} & 1 & 10758.4 & \textbf{7.745e-9} \\
 & \textbf{MadNCL (GPU)} & 0 & - & - \\
 & \textbf{Ipopt  (CPU)} & 1 & 18689.26 & 1.954e-7 \\
 & \textbf{MadNLP (LK, CPU)} & 1 & 15667.51 & 4.728e-8 \\
\hline
\end{tabular}
\caption*{\footnotesize Cvio = maximum constraint violation.}
\end{table}

To address the limitations of \gls*{gpu} solvers at higher tolerance, we also provide the ability to warm start the \gls*{goc3} model using an existing solution. The availability of the full model sheds light on the solution quality of existing results. When running on the unit commitment solutions generated by the benchmark solver (nvar 140k-330k), objective values 0.08\% greater than the benchmark were achieved. The best known solutions were still $\approx$0.005\% better than those found using ExaModelsPower.jl, but generating the \gls*{goc3} model using the best known UC solutions would likely yield new best known solutions to the full problem. Objective quality results were generated separately from the benchmarking results shown earlier.  

\subsection{Break-Down of Total Runtime}
This section analyzes the total runtime time (time between initial call and reported solution) to solve the 78,484 node epigrids ACOPF case.
The total runtime can be  broken down into model construction time, solver initialization time, \gls*{ad} time, linear solve time, and solver internal time.
Model construction time is the time spent outside the solver (e.g., building the JuMP model or ExaModel); solver initialization time is the time spent in the solver before the first iteration (e.g., performing symbolic factorization); \gls*{ad} time is the time spent computing derivatives; linear solve time is the time spent solving linear systems; and solver internal time is the remainder (e.g., computing step directions).
The results in \Cref{sec:num-static,sec:num-mp,sec:num-goc} reports the time excluding model construction time, which is the common practice in solver benchmarking.
When the same problem is solved again with the same sparsity pattern (e.g., no switching), the model construction time and solver initialization time can be eliminated.

\begin{figure}
\centering
\label{time_plot}
\begin{tikzpicture}[font=\scriptsize]
\begin{axis}[
    xlabel = {Solver Configuration (Runtime [s])},
    ylabel = Runtime Proportion,
    width=.85\columnwidth,
    height=5cm,
    ybar stacked,
    bar width = .06\columnwidth,
    grid = both,
    xmajorgrids=false,
    xminorgrids=false,
    xtick = data,
    xtick style={draw=none},
    xticklabels={
        {PowerModels-CPU\\(918.5s)},
        {ExaModelsPower-CPU\\(393.5s)},
        {ExaModelsPower-GPU\\(16.1s)},
    },
    xticklabel style={align=center},
    % xticklabels from table={data/time_cumulative.txt}{Network},
    % xmin = 0, xmax= 4785,
    % ymin= 0, ymax= 4785,
    % xmin = 0, xmax= 2600,
    ymin= 0, ymax= 1,
    legend style={
        cells={anchor=west},
        legend pos=outer north east,
        font=\tiny
        }, 
    reverse legend,
]
    % linear solves
    \addplot[blue, fill=blue!40] table [y=lin, meta = method, x expr=\coordindex, col sep=space] {data/time_cumulative.txt} ;
    \addlegendentry{Linear Solve}
    % AD
    \addplot[red, pattern=north east lines, pattern color = red] table [y=ad, meta = method, x expr=\coordindex, col sep=space] {data/time_cumulative.txt} ;
    \addlegendentry{AD}
    % Internal
    \addplot[violet, pattern=crosshatch dots, pattern color=violet] table [y=internal, meta = method, x expr=\coordindex, col sep=space] {data/time_cumulative.txt} ;
    \addlegendentry{Internal}
    % Init
    \addplot[teal, pattern=grid, pattern color=teal] table [y=init, meta = method, x expr=\coordindex, col sep=space] {data/time_cumulative.txt} ;
    \addlegendentry{Initialization}
    % Construction
    \addplot[Orange, pattern=north west lines, pattern color = Orange] table [y=construction, meta = method, x expr=\coordindex, col sep=space] {data/time_cumulative.txt} ;
    \addlegendentry{Construction}
    % -- legend
    % \legend{
    % Final AC,
    % SOC Iterations,
    % UC Iterations
    % }   
    \end{axis}
\end{tikzpicture}
\caption{Proportion of runtime used by model construction, initialization, solver internals, AD, and linear solves on the 78,484 node pglib epigrids case.}
\label{fig:time_plot}
\end{figure}
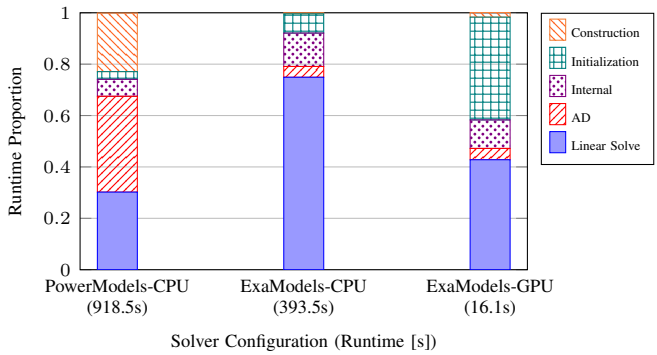

\Cref{fig:time_plot} compares the breakdown of the total runtime of PowerModels.jl.jl and ExaModelsPower.jl on CPU and GPU for solving the 78,484 node pglib epigrids case in polar coordinates at $10^{-4}$ tolerance, using MadNLP.jl as the optimization solver.
PowerModels.jl devotes a significant portion of time to model construction and AD.
Switching to ExaModels.jl, even on CPU, dramatically reduces the time spent on model construction and AD, making linear solve time the main bottleneck.
Solving the problem with ExaModels.jl, MadNLP.jl, and a \gls*{gpu} backend reduces the total time by over an order of magnitude, further reducing the bottleneck from linear solve time.
On GPU solvers, the symbolic analysis phase (included in the initialization time) becomes a significant portion of the total time, as it is not parallelized on GPU.

\setlength{\textfloatsep}{5pt}

%%% Local Variables:
%%% mode: LaTeX
%%% TeX-master: "main"
%%% End:

\section{Conclusions}\label{sec:conc}
This paper introduces ExaModelsPower.jl, a \gls*{gpu}-enabled open-source modeling library for power system optimization.
This library provides a model implementation for various power system optimization problems, focusing on large \gls*{nlp} formulations, such as multi-period \gls*{opf} with and without storage, and the \gls*{goc3} security-constrained formulation.
Our benchmarking results demonstrate that when solved up to moderate precision, \gls*{gpu} solvers can achieve significant speedups compared to CPU solvers, especially for large-scale problems.
This highlights the potential of \gls*{gpu}-accelerated optimization for addressing complex power system problems that were previously computationally prohibitive.
However, none of the tested solvers could solve all instances of the \gls*{goc3} formulation up to $10^{-8}$ precision.
This underscores the need for further research into alternative problem formulations that enable practical solution times for security-constrained multi-period \gls*{ac}\gls*{opf} at scale and understanding the tradeoffs between such formulations. Future work can also augment ExaModelsPower.jl to solve numerous ACOPF problems in batch.

\section*{Acknowledgments}
This material is based upon work supported by the National Science Foundation Graduate Research Fellowship under Grant No. DGE-2141064.
We acknowledge Archim Jhunjhunwala for implementing ExaPowerIO.jl, which provides the efficient data loading capabilities used in ExaModelsPower.jl.

\bibliographystyle{ieeetr} 
\bibliography{PSCC26,shin}

\begin{thebibliography}{10}

\bibitem{cain_history_2012}
M.~B. Cain, R.~P. O’Neill, and A.~Castillo, ``History of {Optimal} {Power}
  {Flow} and {Formulations},'' 2012.

\bibitem{frank_introduction_2016}
S.~Frank and S.~Rebennack, ``An introduction to optimal power flow: {Theory},
  formulation, and examples,'' {\em IIE Transactions}, vol.~48, pp.~1172--1197,
  Dec. 2016.
\newblock Publisher: Taylor \& Francis \_eprint:
  https://doi.org/10.1080/0740817X.2016.1189626.

\bibitem{coffrin_powermodels_2018}
C.~Coffrin, R.~Bent, K.~Sundar, Y.~Ng, and M.~Lubin, ``{PowerModels.jl}: An
  open-source framework for exploring power flow formulations,'' in {\em 2018
  {Power} {Systems} {Computation} {Conference} ({PSCC})}, pp.~1--8, June 2018.

\bibitem{oneill_recent_2011}
R.~P. O’Neill, T.~Dautel, and E.~Krall, ``Recent {ISO} software enhancements
  and future software and modeling plans,'' {\em Federal Energy Regulatory
  Commission, Tech. Rep}, 2011.

\bibitem{aravena_recent_2023}
I.~Aravena, D.~K. Molzahn, S.~Zhang, C.~G. Petra, F.~E. Curtis, S.~Tu,
  A.~Wächter, E.~Wei, E.~Wong, A.~Gholami, K.~Sun, X.~A. Sun, S.~T. Elbert,
  J.~T. Holzer, and A.~Veeramany, ``Recent {Developments} in
  {Security}-{Constrained} {AC} {Optimal} {Power} {Flow}: {Overview} of
  {Challenge} 1 in the {ARPA}-{E} {Grid} {Optimization} {Competition},'' {\em
  Operations Research}, vol.~71, pp.~1997--2014, Nov. 2023.
\newblock Publisher: INFORMS.

\bibitem{holzer_grid_2024}
J.~T. Holzer, C.~J. Coffrin, C.~DeMarco, R.~Duthu, S.~T. Elbert, B.~C.
  Eldridge, T.~Elgindy, M.~Garcia, S.~L. Greene, N.~Guo, and {others}, ``Grid
  optimization competition challenge 3 problem formulation,'' tech. rep.,
  Pacific Northwest National Laboratory (PNNL), Richland, WA (United States),
  2024.

\bibitem{holzer2025go}
J.~T. Holzer, S.~Elbert, H.~Mittelmann, R.~O’Neill, and H.~Oh, ``{GO
  Competition Challenge 3: Problem, solvers, and solution analysis},'' {\em
  Energy Systems}, 2025.

\bibitem{elbert2024final_report}
S.~T. Elbert, J.~T. Holzer, A.~Veeramany, C.~DeMarco, H.~Mittelmann, and
  T.~Overbye, ``{Supporting ARPA-E Power Grid Optimization (Final Report)},''
  tech. rep., Pacific Northwest National Laboratory, 2024.

\bibitem{brun2025alternating}
M.~Brun, T.~Lee, D.~Lauinger, X.~Chen, and X.~A. Sun, ``Alternating methods for
  large-scale ac optimal power flow with unit commitment,'' 2025.

\bibitem{chevalier2024parallelized}
S.~Chevalier, ``{A parallelized, Adam-based solver for reserve and security
  constrained AC unit commitment},'' {\em Electric Power Systems Research},
  vol.~235, p.~110685, 2024.

\bibitem{sharadga2025scalable}
H.~Sharadga, J.~Mohammadi, C.~Crozier, and K.~Baker, ``Scalable solutions for
  security-constrained optimal power flow with multiple time steps,'' {\em IEEE
  Transactions on Industry Applications}, 2025.

\bibitem{geth_flexible_2020}
F.~Geth, C.~Coffrin, and D.~Fobes, ``A flexible storage model for power network
  optimization,'' in {\em Proceedings of the {Eleventh} {ACM} {International}
  {Conference} on {Future} {Energy} {Systems}}, e-{Energy} '20, (New York, NY,
  USA), pp.~503--508, Association for Computing Machinery, June 2020.

\bibitem{petra_solving_2021}
C.~G. Petra and I.~Aravena, ``Solving realistic security-constrained optimal
  power flow problems,'' Oct. 2021.
\newblock arXiv:2110.01669 [math].

\bibitem{shin_accelerating_2024}
S.~Shin, M.~Anitescu, and F.~Pacaud, ``Accelerating optimal power flow with
  {GPUs}: {SIMD} abstraction of nonlinear programs and condensed-space
  interior-point methods,'' {\em Electric Power Systems Research}, vol.~236,
  p.~110651, Nov. 2024.

\bibitem{zimmerman_matpower_2011}
R.~D. Zimmerman, C.~E. Murillo-Sánchez, and R.~J. Thomas, ``{MATPOWER}:
  {Steady}-{State} {Operations}, {Planning}, and {Analysis} {Tools} for {Power}
  {Systems} {Research} and {Education},'' {\em IEEE Transactions on Power
  Systems}, vol.~26, pp.~12--19, Feb. 2011.

\bibitem{orban_nlpmodelsjl_2023}
D.~Orban, A.~S. Siqueira, and {contributors}, ``{NLPModels}.jl: {Data}
  {Structures} for {Optimization} {Models},'' Mar. 2023.

\bibitem{bonaldo_genx_2025}
L.~Bonaldo, S.~Chakrabarti, F.~Cheng, Y.~Ding, J.~Jenkins, Q.~Luo,
  R.~Macdonald, D.~Mallapragada, A.~Manocha, G.~Mantegna, J.~Morris,
  N.~Patankar, F.~Pecci, A.~Schwartz, J.~Schwartz, G.~Schivley, N.~Sepulveda,
  and Q.~Xu, ``{GenX},'' July 2025.

\bibitem{ho_regional_2021}
J.~Ho, J.~Becker, M.~Brown, P.~Brown, I.~Chernyakhovskiy, S.~Cohen, W.~Cole,
  S.~Corcoran, K.~Eurek, W.~Frazier, P.~Gagnon, N.~Gates, D.~Greer, P.~Jadun,
  S.~Khanal, S.~Machen, M.~Macmillan, T.~Mai, M.~Mowers, C.~Murphy, A.~Rose,
  A.~Schleifer, B.~Sergi, D.~Steinberg, Y.~Sun, and E.~Zhou, ``Regional
  {Energy} {Deployment} {System} ({ReEDS}) {Model} {Documentation}: {Version}
  2020,'' {\em Renewable Energy}, 2021.

\bibitem{brown_pypsa_2018}
T.~Brown, J.~Hörsch, and D.~Schlachtberger, ``{PyPSA}: {Python} for {Power}
  {System} {Analysis} {\textbar} {Journal} of {Open} {Research} {Software},''
  Jan. 2018.

\bibitem{c_papadopoulos_r_johnson_and_f_valdebenito_plexos_integrated_plexos_2014}
{C Papadopoulos, R Johnson, and F Valdebenito. PLEXOS® Integrated},
  ``{PLEXOS}® {Integrated} {Energy} {Modelling} around the {Globe},'' 2014.

\bibitem{santos_xavier_unitcommitmentjl_2024}
A.~Santos~Xavier, A.~M. Kazachkov, O.~Yurdakul, J.~He, and F.~Qiu,
  ``{UnitCommitment}.jl: {A} {Julia}/{JuMP} {Optimization} {Package} for
  {Security}-{Constrained} {Unit} {Commitment},'' May 2024.

\bibitem{lubin_jump_2023}
M.~Lubin, O.~Dowson, J.~D. Garcia, J.~Huchette, B.~Legat, and J.~P. Vielma,
  ``{JuMP} 1.0: recent improvements to a modeling language for mathematical
  optimization,'' {\em Mathematical Programming Computation}, vol.~15,
  pp.~581--589, Sept. 2023.

\bibitem{lincoln_pypower_2025}
R.~Lincoln, ``{PYPOWER},'' Sept. 2025.
\newblock original-date: 2009-05-22T12:14:47Z.

\bibitem{thurner_pandapoweropen-source_2018}
L.~Thurner, A.~Scheidler, F.~Schäfer, J.-H. Menke, J.~Dollichon, F.~Meier,
  S.~Meinecke, and M.~Braun, ``Pandapower—{An} {Open}-{Source} {Python}
  {Tool} for {Convenient} {Modeling}, {Analysis}, and {Optimization} of
  {Electric} {Power} {Systems},'' {\em IEEE Transactions on Power Systems},
  vol.~33, pp.~6510--6521, Nov. 2018.

\bibitem{noauthor_exago_nodate}
``{ExaGO}: {High}-performance power grid optimization for stochastic,
  security-constrained, and multi-period {ACOPF} problems..''

\bibitem{balay_petsctao_2025}
S.~Balay, S.~Abhyankar, M.~F. Adams, S.~Benson, J.~Brown, P.~Brune,
  K.~Buschelman, E.~M. Constantinescu, L.~Dalcin, A.~Dener, V.~Eijkhout,
  J.~Faibussowitsch, W.~D. Gropp, V.~Hapla, T.~Isaac, P.~Jolivet, D.~Karpeev,
  D.~Kaushik, M.~G. Knepley, F.~Kong, S.~Kruger, D.~A. May, L.~C. McInnes,
  R.~T. Mills, L.~Mitchell, T.~Munson, J.~E. Roman, K.~Rupp, P.~Sanan,
  J.~Sarich, B.~F. Smith, H.~Suh, S.~Zampini, H.~Zhang, and J.~Zhang,
  ``{PETSc}/{TAO} {Users} {Manual} {Revision} 3.23,'' Tech. Rep. ANL--21-39 Rev
  3.23, Argonne National Laboratory (ANL), Argonne, IL (United States), Mar.
  2025.

\bibitem{beckingsale_raja_2019}
D.~A. Beckingsale, J.~Burmark, R.~Hornung, H.~Jones, W.~Killian, A.~J. Kunen,
  O.~Pearce, P.~Robinson, B.~S. Ryujin, and T.~R. Scogland, ``{RAJA}:
  {Portable} {Performance} for {Large}-{Scale} {Scientific} {Applications},''
  in {\em 2019 {IEEE}/{ACM} {International} {Workshop} on {Performance},
  {Portability} and {Productivity} in {HPC} ({P3HPC})}, pp.~71--81, Nov. 2019.

\bibitem{beckingsale_umpire_2020}
D.~A. Beckingsale, M.~J. McFadden, J.~P.~S. Dahm, R.~Pankajakshan, and R.~D.
  Hornung, ``Umpire: {Application}-focused management and coordination of
  complex hierarchical memory,'' {\em IBM Journal of Research and Development},
  vol.~64, pp.~00:1--00:10, May 2020.

\bibitem{petra_hiop_2018}
C.~G. Petra, N.~Chiang, and J.~Wang, ``{HiOp} – {User} {Guide},'' Tech. Rep.
  LLNL-SM-743591, Center for Applied Scientific Computing, Lawrence Livermore
  National Laboratory, 2018.

\bibitem{khaloie2025review}
H.~Khaloie, M.~Dolanyi, J.-F. Toubeau, and F.~Vall{\'e}e, ``Review of machine
  learning techniques for optimal power flow,'' {\em Applied Energy}, vol.~388,
  p.~125637, 2025.

\bibitem{donti2021dc3}
P.~L. Donti, D.~Rolnick, and J.~Z. Kolter, ``Dc3: A learning method for
  optimization with hard constraints,'' {\em arXiv preprint arXiv:2104.12225},
  2021.

\bibitem{fioretto2020predicting}
F.~Fioretto, T.~W. Mak, and P.~Van~Hentenryck, ``Predicting {AC} optimal power
  flows: Combining deep learning and {L}agrangian dual methods,'' in {\em
  Proceedings of the AAAI conference on artificial intelligence}, vol.~34,
  pp.~630--637, 2020.

\bibitem{pilotoCANOSFastScalable2024}
L.~Piloto, S.~Liguori, S.~Madjiheurem, M.~Zgubic, S.~Lovett, H.~Tomlinson,
  S.~Elster, C.~Apps, and S.~Witherspoon, ``{{CANOS}}: {{A Fast}} and
  {{Scalable Neural AC-OPF Solver Robust To N-1 Perturbations}},'' Mar. 2024.

\bibitem{babaeinejadsarookolaee_power_2021}
S.~Babaeinejadsarookolaee, A.~Birchfield, R.~D. Christie, C.~Coffrin,
  C.~DeMarco, R.~Diao, M.~Ferris, S.~Fliscounakis, S.~Greene, R.~Huang,
  C.~Josz, R.~Korab, B.~Lesieutre, J.~Maeght, T.~W.~K. Mak, D.~K. Molzahn,
  T.~J. Overbye, P.~Panciatici, B.~Park, J.~Snodgrass, A.~Tbaileh, P.~V.
  Hentenryck, and R.~Zimmerman, ``The {Power} {Grid} {Library} for
  {Benchmarking} {AC} {Optimal} {Power} {Flow} {Algorithms},'' Jan. 2021.
\newblock arXiv:1908.02788 [math].

\bibitem{shin_scalable_2024}
S.~Shin, V.~Rao, M.~Schanen, D.~A. Maldonado, and M.~Anitescu, ``Scalable
  {Multi}-{Period} {AC} {Optimal} {Power} {Flow} {Utilizing} {GPUs} with {High}
  {Memory} {Capacities},'' in {\em 2024 {Open} {Source} {Modelling} and
  {Simulation} of {Energy} {Systems} ({OSMSES})}, pp.~1--6, Sept. 2024.

\bibitem{regev_hykkt_2023}
S.~Regev, N.-Y. Chiang, E.~Darve, C.~G. Petra, M.~A. Saunders, K.~Świrydowicz,
  and S.~Peleš, ``{HyKKT}: a hybrid direct-iterative method for solving {KKT}
  linear systems,'' {\em Optimization Methods and Software}, vol.~38,
  pp.~332--355, Mar. 2023.
\newblock Publisher: Taylor \& Francis \_eprint:
  https://doi.org/10.1080/10556788.2022.2124990.

\bibitem{hartPyomoModelingSolving2011}
W.~E. Hart, J.-P. Watson, and D.~L. Woodruff, ``Pyomo: Modeling and solving
  mathematical programs in {{Python}},'' {\em Mathematical Programming
  Computation}, vol.~3, pp.~219--260, Sept. 2011.

\bibitem{anderssonCasADiSoftwareFramework2019}
J.~A.~E. Andersson, J.~Gillis, G.~Horn, J.~B. Rawlings, and M.~Diehl,
  ``{{CasADi}}: A software framework for nonlinear optimization and optimal
  control,'' {\em Mathematical Programming Computation}, vol.~11, pp.~1--36,
  Mar. 2019.

\bibitem{fourerAMPLMathematicalProgramming}
R.~Fourer, ``{{AMPL}}: {{A Mathematical Programming Language}},''

\bibitem{wachter_implementation_2006}
A.~Wächter and L.~T. Biegler, ``On the implementation of an interior-point
  filter line-search algorithm for large-scale nonlinear programming,'' {\em
  Mathematical Programming}, vol.~106, pp.~25--57, Mar. 2006.

\bibitem{pardalos_knitro_2006}
R.~H. Byrd, J.~Nocedal, and R.~A. Waltz, ``Knitro: {An} {Integrated} {Package}
  for {Nonlinear} {Optimization},'' in {\em Large-{Scale} {Nonlinear}
  {Optimization}} (P.~Pardalos, G.~Di~Pillo, and M.~Roma, eds.), vol.~83,
  pp.~35--59, Boston, MA: Springer US, 2006.
\newblock Series Title: Nonconvex Optimization and Its Applications.

\bibitem{duffMA57aCodeSolution2004}
I.~S. Duff, ``{{MA57---a}} code for the solution of sparse symmetric definite
  and indefinite systems,'' {\em ACM Trans. Math. Softw.}, vol.~30,
  pp.~118--144, June 2004.

\bibitem{schenkSolvingUnsymmetricSparse2004}
O.~Schenk and K.~G{\"a}rtner, ``Solving unsymmetric sparse systems of linear
  equations with {{PARDISO}},'' {\em Future Generation Computer Systems},
  vol.~20, pp.~475--487, Apr. 2004.

\bibitem{bezansonJuliaFreshApproach2017}
J.~Bezanson, A.~Edelman, S.~Karpinski, and V.~B. Shah, ``Julia: {{A Fresh
  Approach}} to {{Numerical Computing}},'' {\em SIAM Review}, vol.~59,
  pp.~65--98, Jan. 2017.

\bibitem{nurkanovic2023solving}
A.~Nurkanović, A.~Pozharskiy, and M.~Diehl, ``Solving mathematical programs
  with complementarity constraints arising in nonsmooth optimal control,'' {\em
  Vietnam Journal of Mathematics}, vol.~53, p.~659–697, Aug. 2024.

\bibitem{stephen_elbert_arpa-e_2024}
S.~Elbert, J.~Holzer, A.~Veeramany, R.~O'Neill, H.~Mittelmann, and C.~Coffrin,
  ``{ARPA}-{E} {Grid} {Optimization} ({GO}) {Competition} {Challenge} 3,''
  2024.

\bibitem{tasseff_exploring_2019}
B.~Tasseff, C.~Coffrin, A.~Wachter, and C.~Laird, ``Exploring benefits of
  linear solver parallelism on modern nonlinear optimization applications,''
  {\em arXiv: Optimization and Control}, Sept. 2019.

\bibitem{pacaud2024condensed}
F.~Pacaud, S.~Shin, A.~Montoison, M.~Schanen, and M.~Anitescu,
  ``Condensed-space methods for nonlinear programming on gpus,'' {\em arXiv
  preprint arXiv:2405.14236}, 2024.

\bibitem{montoison_madncl_2025}
A.~Montoison, F.~Pacaud, M.~Saunders, S.~Shin, and D.~Orban, ``{MadNCL}: {A}
  {GPU} {Implementation} of {Algorithm} {NCL} for {Large}-{Scale}, {Degenerate}
  {Nonlinear} {Programs},'' Oct. 2025.
\newblock arXiv:2510.05885 [math].

\bibitem{grigg_ieee_1999}
C.~Grigg, P.~Wong, P.~Albrecht, R.~Allan, M.~Bhavaraju, R.~Billinton, Q.~Chen,
  C.~Fong, S.~Haddad, S.~Kuruganty, W.~Li, R.~Mukerji, D.~Patton, N.~Rau,
  D.~Reppen, A.~Schneider, M.~Shahidehpour, and C.~Singh, ``The {IEEE}
  {Reliability} {Test} {System}-1996. {A} report prepared by the {Reliability}
  {Test} {System} {Task} {Force} of the {Application} of {Probability}
  {Methods} {Subcommittee},'' {\em IEEE Transactions on Power Systems},
  vol.~14, pp.~1010--1020, Aug. 1999.

\bibitem{department_of_market_monitoring_2024_2025}
D.~of~Market~Monitoring, ``2024 {Special} {Report} on {Battery} {Storage},''
  tech. rep., California ISO, 2025.

\bibitem{california_iso_gross_2023}
C.~ISO, ``Gross \& net load peaks fact sheet,'' tech. rep., 2023.

\bibitem{parker_managing_2024}
R.~Parker and C.~Coffrin, ``Managing power balance and reserve feasibility in
  the {AC} unit commitment problem,'' {\em Electric Power Systems Research},
  vol.~234, p.~110670, Sept. 2024.

\end{thebibliography}

\end{document}